\documentclass{aa}
\usepackage{times}
\usepackage{amssymb}
\usepackage{epsfig}

\title{Constraints on $(\Omega_{\rm m},\Omega_\Lambda)$ using distributions of
inclination angles for high redshift filaments.}

\newcommand{\Lya}{Ly$\alpha$}

\author{M.~Weidinger\inst{1} \and P.~M\o ller\inst{2}  \and
J.P.U.~Fynbo\inst{2} \and B.~Thomsen\inst{1} \and M.P.~Egholm\inst{1}}

\institute{ University of Aarhus, Ny Munkegade,
            DK-8000 \AA rhus C,
            Denmark
\and
            European Southern Observatory,
            Karl-Schwarzschild-Stra\ss e 2,
            D-85748, Garching by M\"unchen,
            Germany
          }
\offprints{M. Weidinger}
\mail{michaelw@ifa.au.dk}
\date{Received / Accepted }

\begin{document}

\abstract{In this paper we present a scale free method to determine
the cosmological parameters $(\Omega_{\rm m},\Omega_\Lambda)$. The
method is based on the requirement of isotropy of the distribution
of orientations of cosmological filaments.
The current structure formation paradigm predicts that the
first structures to form are voids and filaments, causing a
web-like structure of the matter distribution at high redshifts.
Recent observational evidence suggests that the threads, or filaments,
of the cosmic web most easily are mapped in Ly$\alpha$ emission. We
describe how such a 3D map can be used to constrain the cosmological
parameters in a way which, contrary to most other cosmological tests,
does not require the use of a standard rod or a standard candle.
We perform detailed simulations in order to define the optimal
survey parameters for the definition of an observing programme aimed
to address this test, and to investigate how statistical and
observational errors will influence the results. We conclude that
observations should target filaments of comoving size $15-50$ Mpc in
the redshift range $2-4$, and that each filament must be defined by
at least four Ly$\alpha$ emitters. Detection of 20 filaments will be
sufficient to obtain a result, while 50 filaments
will make it possible to place significant new constraints on
the values of $\Omega_{\rm m}$ and $\Omega_\Lambda$ permitted by the
current supernova observations.  In a future paper we study how 
robust these conclusions are to systematic velocities in the survey
box.
\keywords{cosmology: theory --- cosmological parameters}  }

\titlerunning{New constraints on $(\Omega_{\rm m},\Omega_\Lambda)$}
\maketitle

\section{Introduction}
Recent studies have successfully narrowed down the permitted parameter
space for the cosmological parameters $\Omega_{\rm m}$ and
$\Omega_\Lambda$. Confidence intervals defined by observations of
distant supernovae (Riess et al.~1998, Perlmutter et al.~1999) and
those resulting from the high-resolution observations of the cosmic
microwave background radiation (CMB) (Jaffe et al.~2000) meet almost
orthogonally, and
as a result they combine to bracket a domain of high
probability. There remains, however, the question of possible
systematic effects which may produce significant errors (e.g.~Simonsen
\& Hannestad 1999; Rowan-Robinson 2002). It is
therefore important that new and independent methods to determine
$\Omega_{\rm m}$ and $\Omega_\Lambda$ should be sought and exploited.

The classic cosmological tests (Sandage 1961) involve the use of
either standard rods or standard candles. The supernovae projects for
example, are based on the use of standard candles, while
recent proposals to use the distribution of Lyman Break Galaxies and
arrangements of \Lya-forest systems (Roukema 2001 and McDonald \&
Miralda-Escud\'e 1999 respectively) involve the use of standard rods.
The potential worry with all such methods is that the hypothetical
standard rods or candles may in reality not be good standards because of
evolutionary effects, or simply because the evolution of structure
may be a fractal process without any preferred scale.

The simplest and cleanest cosmological tests one might imagine would
be a purely geometric test not involving measurements of standard rods.
The first such method was proposed by Alcock \& Paczy\'nski (1979), who
considered an idealized set of objects distributed spherically symmetric
and on average following the Hubble flow. The idea is beautiful in its
simplicity, but the requirement of spherical symmetry of the
distribution (requiring a length scale to be the same in all
directions) is in reality an indirect use of a standard rod, even if at
a single point in redshift space. The hypothetical spherical
distribution might as well be ellipsoidal, which would then
invalidate the result. The question is if it is at all possible to
devise a geometrical test not requiring any length scale at all.

Numerical simulations of early structure formation based on the Cold
Dark Matter scenario (Klypin \& Shandarin 1983, White et al.~1987,
Rauch et al.~1997) show that high redshift star-forming regions tend
to align themselves in long filaments with a higher-than-average matter
density. This has recently been confirmed observationally by
M{\o}ller \& Fynbo (2001), who also pointed out that such structures
offered three independent cosmological tests, all of which were purely
geometric, and none of which required the early structure formation to
have a preferred length scale. The only requirement is that of global
isotropy. The first of those three
tests concerns the distribution of filament inclination angles which,
in an isotropic universe, must be random. In this paper we present
a detailed analysis, based on Monte Carlo simulations, of this first
filament test. 

The paper is organized as follows: In section 2 we introduce the
method, in section 3 we describe the details of the simulations and
interpret the results in a cosmological context to see which constraints
we may put on the cosmological parameters, in section 4 we address some
additional sources of statistical and observational errors, and use
the results to define the optimal survey parameters for an observing
programme, and finally in section 5 we summarize the results and
conclusions. We have included a short appendix, which briefly
introduces the notation and the necessary cosmological relations.

\section{F({\bf $\Omega_{\rm m}$, $\Omega_\Lambda$}, z): The line of
sight scale factor}\label{method}
The cosmological principle tells us that any class of elongated objects
or structures will, at all redshifts, display a random distribution of
orientations of their major axes. Elongated structures that are so large
that they are everywhere anchored in the local Hubble flow are usually
referred to as `filaments' and `walls', and for such objects the
cosmological redshift will vary along the structures and thereby reveal
their orientations in 3D redshift space. The basis of the cosmological
test described here is to determine the sets of
($\Omega_{\rm m}$, $\Omega_\Lambda$) which, at any given redshift, will
make the observed distribution of orientations anchored in the
Hubble flow conform to this requirement of isotropy.

As a simple example let us first consider an area on the sky in the form
of a square with sides of angular size $\phi$. At any given redshift $z$
this will correspond to a square in the plane of the sky with sides of
proper size
\begin{equation}
W_{\rm true} = \frac{\phi}{H_0}
f_{\rm W}(\Omega_{\rm m}, \Omega_\Lambda, z)
\end{equation}
where prescriptions for calculation of $f_{\rm W}$ (as well as of
$f_{\rm L}$ below) are provided in appendix A. If we further consider a
redshift interval $\delta z$ centred on $z$, we have defined a box of
proper length
\begin{equation}
L_{\rm true} = \frac{\delta z}{H_0}
f_{\rm L}(\Omega_{\rm m}, \Omega_\Lambda, z).
\end{equation}
Let the box be filled with a large number of filaments, and let their
distribution of orientations be isotropic. If one were to either stretch
or squeeze the box along the line of sight (along the redshift
direction), the angular distribution of the filaments would (because
they are anchored in the Hubble-flow) no longer remain isotropic.

For illustration purposes it is useful to think of a box inside which an
isotropic network of rubber bands has been mounted between opposing
walls. If we stretch the box along one direction the angles between the
rubber bands and the stretching direction will all decrease. If,
instead, we were to stretch the box by that same factor in {\it all}
directions, the angles would remain unchanged. Hence, in considering
observationally only the distribution of {\it angles}, one is not
gaining any information about true sizes. By the same token, since the
true size of an object falls out of the equations, the test that we
describe here does not require the existence of a `standard rod' at any
redshift.

Because the actual length and width of the box is irrelevant for our
test, the only numerical value of interest is the ``change of scale''
between the two directions (along the line of sight and perpendicular
to the line of sight). What we are interested in is therefore the ratio
between the length and the width
\begin{equation}
\frac{L}{W} =
\frac{\delta z}{\phi} \frac{f_{\rm L}(\Omega_{\rm m}, \Omega_\Lambda, z)}
{f_{\rm W}(\Omega_{\rm m}, \Omega_\Lambda, z)}
\end{equation}
and how the {\it apparent} value of that ratio relates to its {\it
true} value
\begin{equation}
F(\Omega_{\rm m}, \Omega_\Lambda, z) = \frac{(L/W)_{\rm app}}{(L/W)_{\rm true}}.
\end{equation}
Note that $H_0$, $\phi$, and $\delta z$ have cancelled out of the
ratio $F(\Omega_{\rm m}, \Omega_\Lambda, z)$. In what follows we shall
refer to this ratio as the {\it ``line of sight scale factor''}.

\begin{figure}[ht]
\begin{center}
\epsfig{file=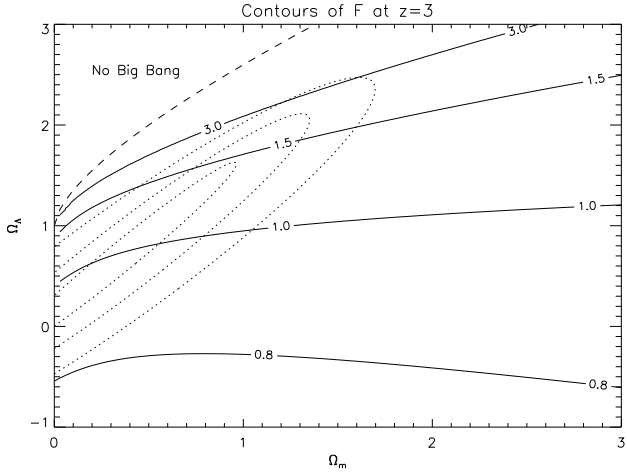,width=9.0cm}
\caption{
Solid: Contours of constant value of the dimensionless line of sight
scale factor $F$ (here for $z=3$ and normalized to $F=1$ for
$\Omega_{\rm m}=0.3$, $\Omega_\Lambda=0.7$).
Dotted: $68.3\%$, $95.4\%$ and $99.7\%$ confidence regions
obtained by supernovae observations (Riess et al.~1998).}\label{fig:Fplot}
\end{center}
\end{figure}

Returning briefly to the example above, for a given triple of
observables $(\phi , z, \delta z)$ and an assumed cosmology
$(\Omega_{\rm m}, \Omega_\Lambda)$, the box is fully defined.
If one now assumes a different cosmology, the inferred size of the
box will change. For most choices of cosmology, the change in length
and width of the box will be different (hence the value of $F$ will
change), but it is always possible to find sets of $(\Omega_{\rm m},
\Omega_\Lambda)$ with the combined effect of changing both length and
width by the same factor, thereby keeping $F$ constant. In
Fig.~\ref{fig:Fplot} we show a set of such ``isoscale curves''
(curves of constant $F$) for a redshift of 3. Our test can now be
formulated simply: ``Determine the sets of $(\Omega_{\rm m},
\Omega_\Lambda)$ for which $F=1$''. The remainder of this paper is
concerned with the formulation of how this is done in practice. In
addition we use Monte Carlo simulated data to determine what level
of accuracy may be reached with current telescopes and instrumentation.

\section{Limits on $F$ from Monte Carlo simulations}\label{simulation}
Consider a large number of idealized, randomly aligned, linear and thin
filaments. Let the inclination angle $\theta$ of a filament be the angle
between the filament major axis and our line of sight to the filament. A
filament pointing towards us will hence have inclination angle zero,
and the cumulative distribution of isotropically distributed
inclination angles can be shown to be
\begin{equation}
P(x) = 1- \frac{1}{\sqrt{1+\tan^2 x}}.
\end{equation}
where $x\in \left[0,\pi/2\right]$ and $P(x)$ is the probability of
observing an inclination angle $\theta < x$. If we now use a wrong set
of values $(\Omega_{\rm m}, \Omega_\Lambda)$, then the {\it apparent}
distribution is no longer isotropic and the accumulated distribution of
inclination angles is instead given by the general expression
\begin{equation}
P(x,F) = 1- \frac{1}{\sqrt{1+F^2\tan^2 x}}.
\end{equation}
where $F$ is the line of sight scale factor (Eq.~4).

Given a set of observed inclination angles, a K-S test (Press et
al.~1989) against $P(x,F)$, can be used to determine the value of $F$
which best fits the observed distribution. At present there is no data
set available large enough to place useful constraints on the
cosmological parameters. Instead we have created Monte Carlo data in
order to determine how large a data set we shall need to be able to set
interesting limits. The K-S test is only valid for fairly large samples,
so in addition we have tested a number of alternative estimators all
based on the comparison of cumulative distributions. The classic K-S
test only seeks to minimize the largest difference between
the two cumulative distributions. For small samples we found that
the scatter of the fit was smaller when we minimized the sum of the
square of the difference at {\it each} entry in the sample (Eq.~7),
while for large samples the two methods gave identical scatter.

The samples of simulated filaments were computed by randomly selecting
$N$ inclination angles, $\theta_i$, such that the values of
$\cos\theta_i$ formed a uniform distribution between $0$ and $1$.
For each such sample we used the extended K-S test described above
and determined the best fitting value of $F$ by minimization of $D(F)$:
\begin{equation}
D(F) = \sum_{i=1}^Nd_i^2,
\end{equation}
where $d_i =\max\left(\left|\frac{i}{N} -
P\left(\theta_i,F\right)\right|,\left| \frac{i-1}{N} -
P\left(\theta_i,F\right)\right|\right)$, and $P$ is defined in
Eq.~6.
As an illustration we show in Fig.~2 the cumulative distribution of a
realization of 50 filaments (histogram) and the predicted distribution
for a range of values of $F$ (lines) based on Eq.~6.

\begin{figure}[h]
\begin{center}
\epsfig{file=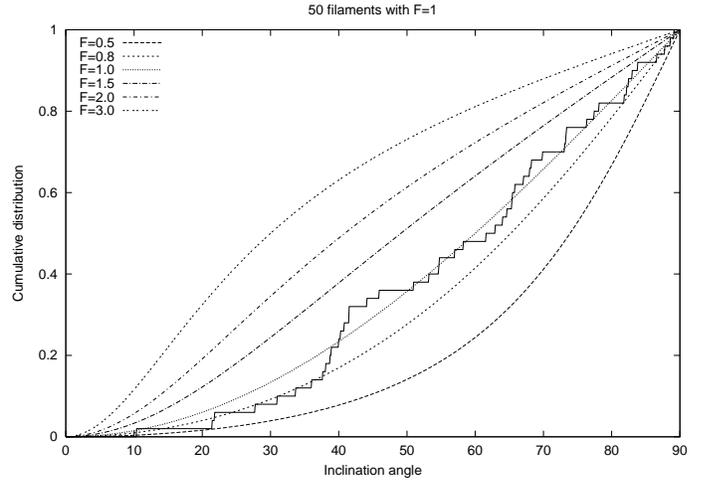,width=6.5cm,angle=-90}
\caption{Cumulative distribution of Monte Carlo simulation of 50
filaments, compared to expected distributions for a range of values
of the line of sight scale factor $F$.}
\label{fig:N50F2}
\end{center}
\end{figure}

\begin{figure}[h]
\begin{center}
\epsfig{file=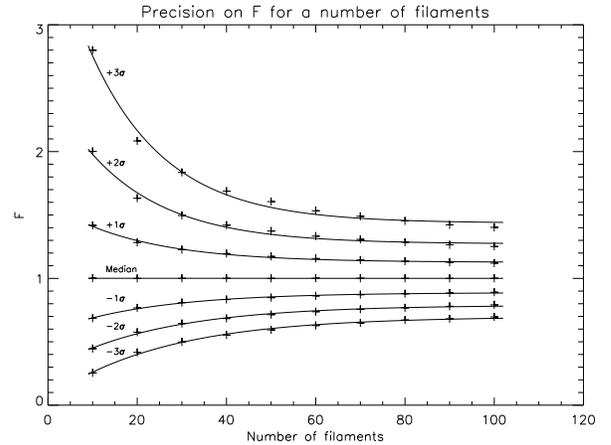,width=8.5cm}
\caption{Plot of the uncertainty with which we can determine the
stretching factor $F$ as a function of how many filaments are in our
sample. This is the result of $2\times 10^4$ simulations for each number
of filaments. The upper and lower 1, 2 and 3$\sigma$ uncertainties are
fitted with exponentials. The result is given in table
\ref{table:fit}.}\label{fig:FN2_fit}
\end{center}
\end{figure}

It is now interesting to ask how many filaments one would need to
observe, to be able to place cosmologically interesting limits on the
observed value of $F$. To address this question we repeated the
simulations for the range 10 to 100 filaments in steps of 10, and in
each case running $2\times 10^4$ simulations. The results are shown in
Fig.~\ref{fig:FN2_fit}. The one-sided 1, 2 and 3$\sigma$ curves were
found as the 15.9\%, 2.28\% and 0.13\% quantiles.  The 1, 2, and
3$\sigma$ curves are fitted well by the exponential functions given in
Table~\ref{table:fit}. The median is consistent with a straight line
with intercept 1.00 and slope 0.000.

\begin{table}[h]
\begin{center}
\caption{The $n\sigma$ errors fitted with
$F(N)=ae^{bN}+c$.}\label{table:fit}
\begin{tabular}{@{}rrrr}
\hline
$n\sigma$ & $a$ & $b$ & $c$\\
\hline
$+3\sigma$ & $2.67$ &  $-0.0650$  & $1.45$\\
$+2\sigma$ & $1.25$ & $-0.0560$ & $1.27$\\
$+1\sigma$ & $0.482$ & $-0.0507$ & $1.13$\\
$-1\sigma$ & $-0.313$ & $-0.0451$ & $0.887$\\
$-2\sigma$ & $-0.498$ & $-0.0396$ & $0.791$\\
$-3\sigma$ & $-0.671$ & $-0.0417$ & $0.694$\\
\hline
\end{tabular}
\end{center}
\end{table}

\begin{figure}[ht]
\begin{center}
\epsfig{file=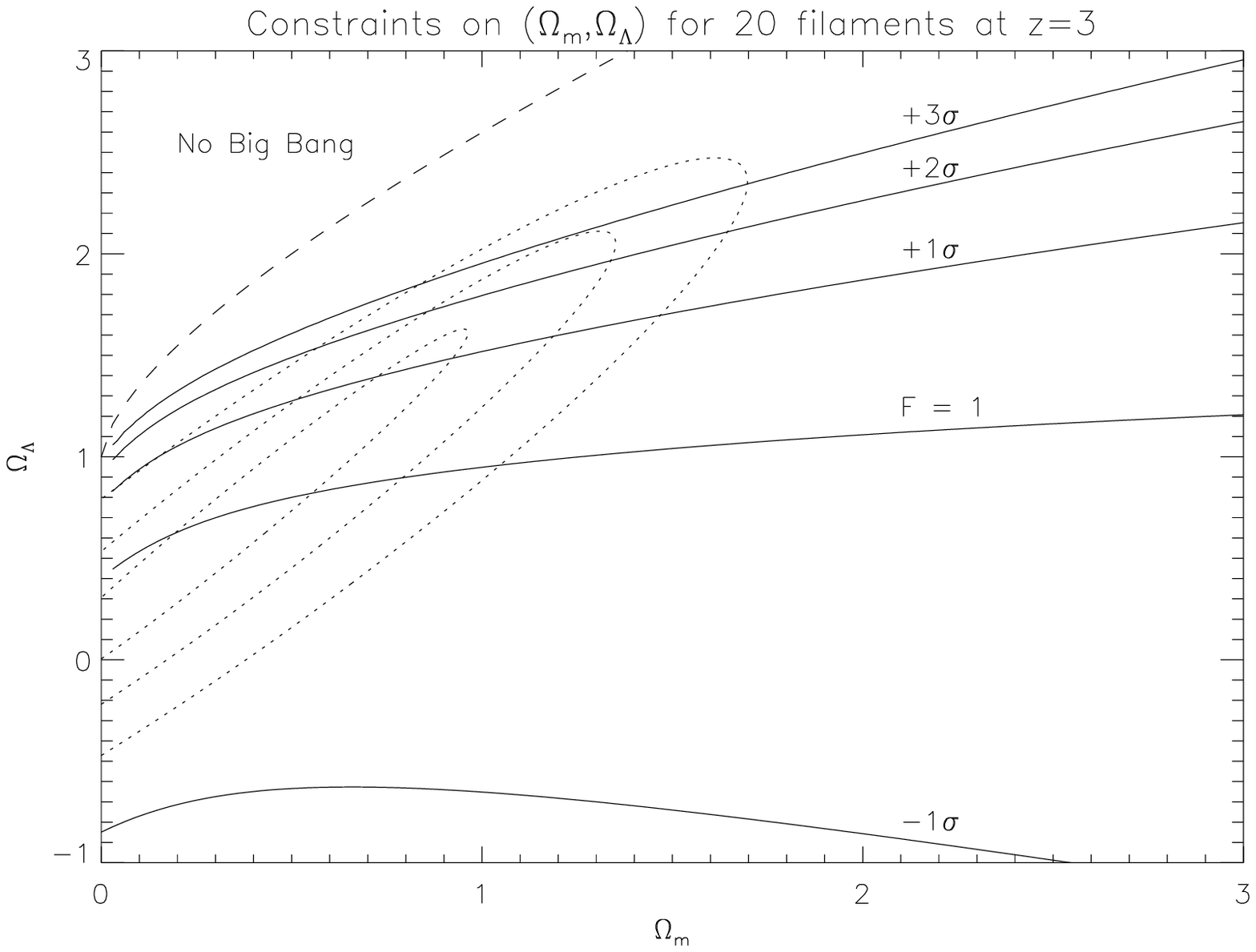,width=8.5cm}
\epsfig{file=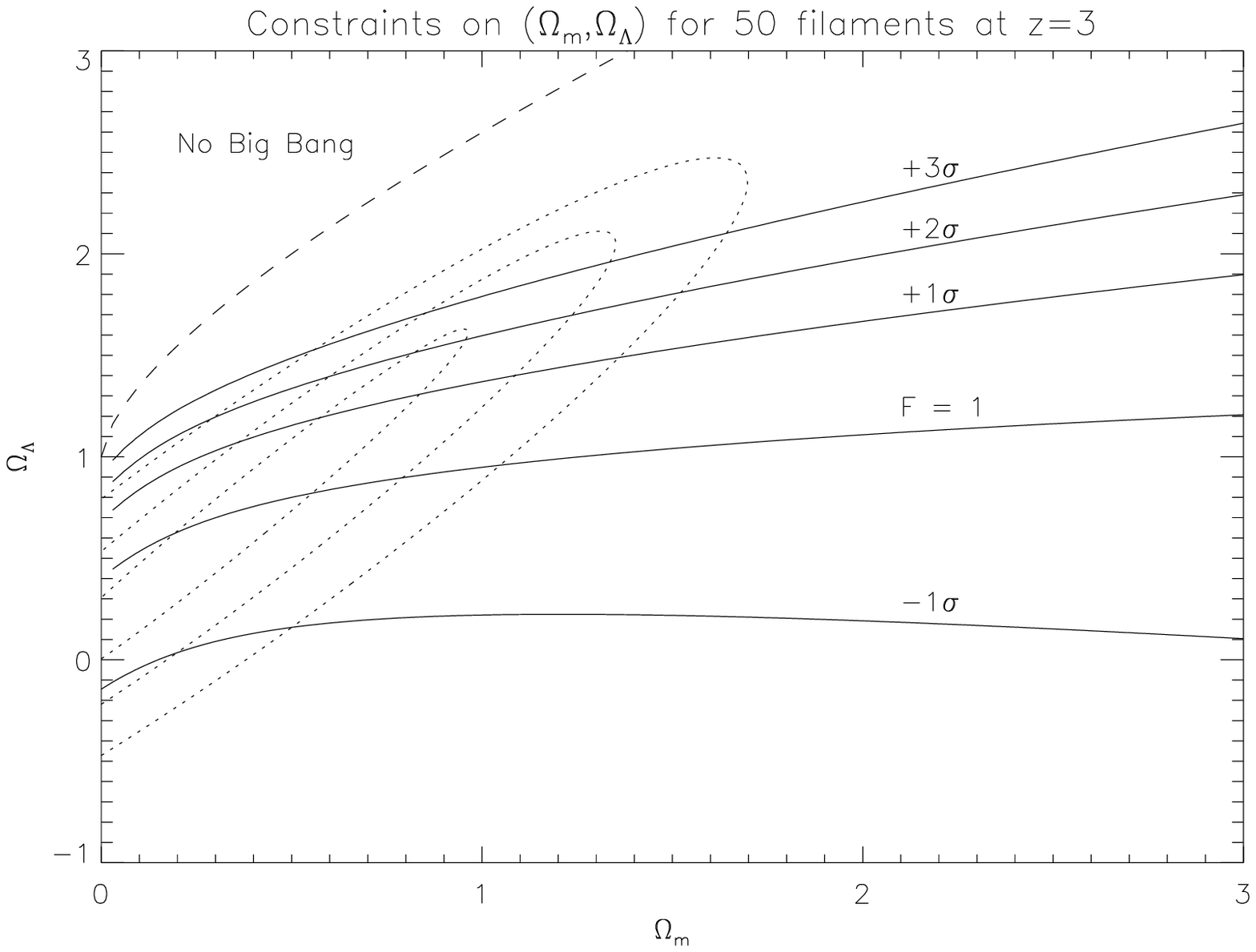,width=8.5cm}
\caption{Solid: Confidence limits on $F$ for observations of 20 (upper
plot) and 50 (lower plot) filaments at $z=3$, normalized to
$\Omega_{\rm m}=0.3$, $\Omega_\Lambda=0.7$. ~~Dotted:
The $68.3\%$ (smallest), $95.4\%$ and $99.7\%$ (largest) confidence
regions obtained by supernovae observations.}\label{fig:N20}
\end{center}
\end{figure}

As illustrated in Fig.~1, each value of $F$ represents, at a given
redshift, a curve in the $(\Omega_{\rm m},\Omega_\Lambda)$
diagram. Using the cosmological relations given in appendix
\ref{sec:relations}, we can therefore transform our limits on $F$
(Fig.~\ref{fig:FN2_fit}) into a set of curves confining the permitted
cosmologies (Fig.~\ref{fig:N20}).  Part of the parameter space in
Fig.~\ref{fig:N20} represents ``bouncing'' universes that do not
expand monotonically from a Big Bang. This corresponds to the region
in the figure where $\Omega_\Lambda > 4\Omega_{\rm m}\:x^3$, in which
\begin{equation}
x = \left\{ \begin{array}{ll}
              \cosh\big[\frac{1}{3}\cosh^{-1}(\Omega_{\rm m}^{-1}-1)\big] &
              \mbox{when $0 < \Omega_{\rm m} \le \frac{1}{2}$} \\
              \cos\big[\frac{1}{3}\cos^{-1}(\Omega_{\rm m}^{-1}-1)\big]   &
              \mbox{when $\frac{1}{2} \le \Omega_{\rm m}$} \end{array}
              \right.
\end{equation}
(Felten \& Isaacman 1986). In this region of the figure, the expression
$\left(1+z\right)^2\left(\Omega_{\rm m}z+1\right) -
\Omega_{\Lambda}z\left(z+2\right)$ under the square-root in the equation
for $\frac{dr}{dz}$ (see Appendix A) becomes negative for some $z$
(the sign of the expression shifts when the universe changes from
expansion to contraction and vice versa). We do not calculate $F$ in
this region.

The contours of $F$ in Fig.~\ref{fig:N20} are calculated for $z=3$, for
20 and 50 filaments, and are normalized to $F=1$ for an
$\Omega_{\rm m}=0.3$, $\Omega_\Lambda=0.7$ universe. For easy comparison,
the confidence regions obtained by supernovae observations
(Riess et al.~1998) are also plotted. Two interesting points are
immediately obvious from Fig.~\ref{fig:N20}. First it is seen that the
isoscale curves are mostly horizontal, and that they place tight upper
limits on $\Omega_\Lambda$ almost independently of $\Omega_{\rm m}$.
This is contrary to the corresponding confidence limits from the
supernovae observations which span diagonal regions in the diagram.
Secondly it is seen that already with a sample of 20 filaments one
would set upper limits on the value of $\Omega_\Lambda$ intersecting
those currently set by the high $z$ SN studies. Such a limited study
would therefore not only provide a new and independent way to determine
$\Omega_\Lambda$, it would already serve to further constrain the
parameter space allowed by the SN studies. For 50 filaments the
intersections would be such that they would reduce the area of the
1$\sigma$ upper confidence limit set by the SN studies by 30\%.

\section{Observational implementation of the test}
\label{results}

\subsection{The optimal redshift range}
\begin{figure}
\epsfig{file=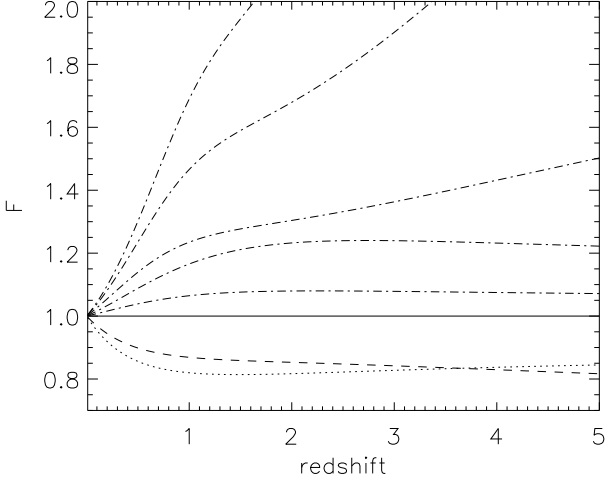, width=8.5cm}
\caption{We plot $F$ as a function of redshift for a range
of cosmological models. $F=1$ corresponds to the $\Omega_{\rm m} = 0.3$,
$\Omega_{\Lambda} = 0.7$ model. The dotted and dashed curves below the
$F=1$ line correspond to $\Omega_{\Lambda} = 0$ models with
$\Omega_{\rm m} = 1$ and $\Omega_{\rm m} = 0.3$ respectively. The
dot-dashed curves above the $F=1$ line correspond to models with
($\Omega_{\rm m}$,$\Omega_{\Lambda}$) = (0.2,0.8),
(0.1,0.9), (0.3,1.2), (0.3,1.4), (0.3,1.5).}
\label{Fz}
\end{figure}

For the planning of an observing campaign to carry out the proposed
test, it is useful first to
consider at which redshifts the test is most sensitive to realistic
values of $\Omega_{\Lambda}$, and how one may best identify filaments
at those redshifts. In Fig.~\ref{Fz} we plot $F$ as a function of
redshift for several different cosmological models, also here
normalized so that $F=1$ corresponds to the $\Omega_{\rm m} = 0.3$,
$\Omega_{\Lambda} = 0.7$ model. The dotted and dashed curves below the
$F=1$ line correspond to $\Omega_{\Lambda} = 0$ models with
$\Omega_{\rm m} = 1$ and $\Omega_{\rm m} = 0.3$ respectively. The
dot-dashed curves above the $F=1$ line correspond to models with
$\Omega_{\Lambda}>0.7$ (see caption for details). All curves converge
at $F=1$ in the limit of zero redshift. It is seen that for all models
with $\Omega_{\Lambda}<1$, $F$ is almost independent of redshift in
the redshift range $z\approx1.5-4$, while $F$ grows monotonically with
redshift for models with $\Omega_{\Lambda}>1$. Hence, in order to test
against any cosmology with $\Omega_{\Lambda}<1$ one should select the
redshift in the range $1.5-4$ where filaments are most easily detected.

\subsection{Filament identification and sparse sampling errors}
Ly$\alpha$ narrow band imaging has proven an efficient technique for
identification of high redshift filaments (M{\o}ller and Warren 1998,
M{\o}ller and Fynbo 2001). Confirming spectroscopy then maps out
filaments as
strings of separate star forming regions, each glowing in Ly$\alpha$.
In what follows we shall refer to those Ly$\alpha$ emitting regions
using the shorter name `LEGOs' (Ly$\alpha$ Emitting Galaxy-building
Objects). Since each filament in this way is mapped by a finite number
of objects, the accuracy with which one may determine its inclination
angle will be a function of the number of LEGOs defining it.

As for the optimal redshift range and number of filaments, it is useful to
consider what may be the minimum number of LEGOs needed per filament for
an adequate determination of its inclination angle. To address this
question we ran a series of simulations in which we randomly placed $N$
LEGOs inside filaments defined as cylinders of length $L$ and diameter
$D$. We then determined the ``observed'' orientation vector of each
filament, (defined as the best fitting straight line through the $N$
points), and the ``observed'' length-to-diameter ratio, $L_{\rm
obs}/D_{\rm obs}$.\footnote{The length of the cylinder is found as $L_{\rm
obs}\equiv \sqrt{12}~{\rm RMS}(z)$, and the diameter $D_{\rm
obs}\equiv\sqrt{8}~{\rm RMS}(r)$, where $z$ is the position of a point
projected onto the axis of symmetry, and $r$ is the distance of a point to
the axis of symmetry.} The observed set of orientation vectors define a
solid angle centred on the true orientation vector, from which the sparse
sampling error ($\sigma_{\rm sp}$) may be determined directly as the RMS
of the 2D distribution of the observed set of intersections between the
orientation vectors and the surface of the unit sphere. In Fig.~6 we show
contours of constant $\sigma_{\rm sp}$ as a function of $N$ and the
$L_{\rm obs}/D_{\rm obs}$ ratio. As one would expect $\sigma_{\rm sp}$ is
small for long/thin filaments but larger for short/wide filaments. It is
also seen that for filaments defined by fewer than four LEGOs $\sigma_{\rm
sp}$ grows rapidly. M{\o}ller \& Fynbo (2001) reported the detection of a
filament at $z=3.04$ defined by eight LEGOs. Assuming $(\Omega_{\rm
m}$,$\Omega_{\Lambda}) = (0.3,0.7)$ and using the procedure outlined
above, we find $L_{\rm obs}=20.2$~Mpc and $D_{\rm obs}=3.2$~Mpc (comoving)
for the $z=3.04$ filament, and an inclination angle of $25.5^\circ$. Its
position is marked in Fig.~6 and it is seen that the sparse sampling error
on its orientation is $\approx 1.9^\circ$.

\begin{figure}[h]
\begin{center}
\epsfig{file=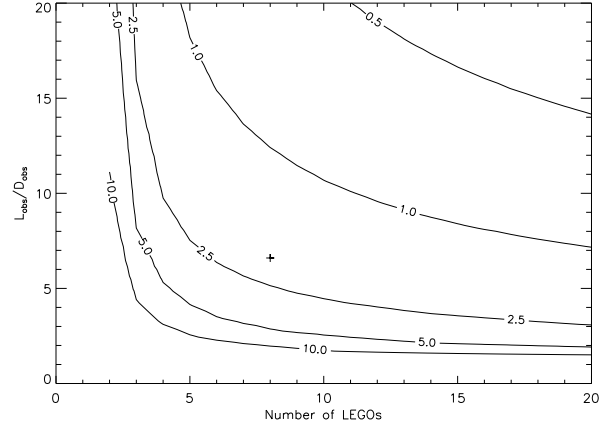,width=8.5cm}
\caption{Contours of constant sparse sampling error on filament
orientation as a function of the number of LEGOs defining
the filament and the length-to-diameter ratio $L_{\rm obs}/D_{\rm obs}$.
The contours are $\sigma_{\rm sp}=0.5^\circ$, $1^\circ$, $2.5^\circ$,
$5^\circ$ and $10^\circ$, and the cross marks the filament detected by
M\o ller \& Fynbo (2001). The contours are based on $10^6$
runs at each point.}
\label{fig:thetasurface}
\end{center}
\end{figure}

In order to propagate the sparse sampling errors shown in Fig.~6 to errors
on the determination of the line of sight scale factor $F$, we repeated
the Monte Carlo simulations of Sect.~3, now including random errors on
filament orientations as described above. The goal was to determine how
large intrinsic errors on the filament orientations we could tolerate
before the error analysis summarised in Fig.~3 would be seriously
compromised. The Monte Carlo simulations were repeated for 20, 50, and 80
filaments, and for $\sigma_{\rm sp}$ in the range $0^\circ$ to $10^\circ$.
The results are summarised in Table~2 where we list errors on $F$ from
Fig.~3, and from the full analysis including sparse sampling errors. It is
seen that even for $\sigma_{\rm sp}$ as large as $10^\circ$ the additional
errors are insignificant. From Fig.~6 we see that $\sigma_{\rm
sp}<10^\circ$ always is achieved for filaments with $L_{\rm obs}/D_{\rm
obs}>3$ defined by four LEGOs or more.

\begin{table}[h]
\begin{center}
\caption{Errors ($1\sigma$) on determination of the line of sight scale
factor $F$ (Fig.~3) including sparse sampling of individual filaments.
The results are based on $2\times 10^5$ runs.
}
\begin{tabular}{c | c c c}
\hline
    &   & Filaments &   \\
$\sigma_{\rm sp}$ &   20 &   50 &   80 \\
\hline
$ 0.0^\circ$          & 0.259 & 0.162 & 0.128 \\
$ 5.0^\circ$          & 0.265 & 0.166 & 0.131 \\
$10.0^\circ$          & 0.283 & 0.179 & 0.140 \\
\hline
\end{tabular}
\label{table:sparsesampling}
\end{center}
\end{table}

In conclusion of this section, a first rough determination of
($\Omega_{\rm m}$, $\Omega_\Lambda$) can be obtained with the discovery of
20 filaments each defined by four LEGOs, or 80 LEGOs in all. With 50
filaments, or 200 LEGOs total, the errors are such that they will place
significant new constraints on the values of $\Omega_{\rm m}$ and
$\Omega_\Lambda$ permitted by the current supernova observations.

\subsection{Optimal survey box size}
Up till this point we have discussed the error budget caused by
purely statistical errors only; errors arising from a finite
filament sample and from the finite number of objects defining each
filament. Those errors are scale invariant and place no constraints on
the size of the survey box. For observational reasons there is an upper
bound to the size of the volume we are able to survey to the necessary
flux limit. On the other hand we can only identify filaments of
sizes smaller than the chosen survey box, and if that box is too small
then measurement errors and errors caused by motion relative to the
Hubble flow could become dominant. One may consider two types of motion
relative to the Hubble flow; ``random motion'' and ``bulk flow''
(streaming).

\subsubsection{Random velocities and measurement errors}
Random velocities and measurement errors are most efficiently treated
together because errors on the redshift are independent from random
radial motion and therefore adds in quadrature. Let us consider the
filament reported by M{\o}ller \& Fynbo (2001). In this
case the error on the redshift determination of the Ly$\alpha$ lines was
50~km~s$^{-1}$ (Fynbo, M{\o}ller \& Thomsen 2001).
At $z\approx 0$ the typical peculiar velocity is 200~km~s$^{-1}$
(Branchini et al.~2001). This includes both random motion and
systematic flows towards large dark matter haloes, and can therefore
be considered a firm upper limit to the random peculiar velocities at
high redshifts. Only the radial component (100~km~s$^{-1}$) of the random
velocity vector is observed, and in combination with the typical
redshift measurement error we have a 112~km~s$^{-1}$ radial error. This
corresponds to $dz=0.0015$ at $z=3$, or to 9\% of the length of the
filament (390~kpc in an $\Omega_{\rm m}=0.3$, $\Omega_\Lambda=0.7$,
$h=0.65$
universe). To understand the effects on a survey for filaments, let us
consider the two limiting cases of filaments with $\theta = 0^\circ$
and $\theta = 90^\circ$. For $\theta = 0^\circ$ the filament is aligned
along our line of sight, and the additional scatter (RMS~=~9\% of
total length) along the filament will simply make it look at bit
longer than it is in reality. Hence the $L_{\rm
obs}/D_{\rm obs}$ ratio will be overestimated by a miniscule amount, and
$\sigma_{\rm sp}$ will be similarly underestimated. Since $\sigma_{\rm
sp}$ is already irrelevantly small, this will have no consequences for
the cosmological test. Considering now the case $\theta = 90^\circ$
(a filament perpendicular to the line of sight) we see that in this
case the measured diameter of the filament (800~kpc) will be
increased (along the line of sight) by the added scatter, resulting in
an added error on the determination of the inclination angle. The
added error can be found from
Fig.~6 via calculation of the predicted $L_{\rm obs}/D_{\rm obs}$ ratio.
The filament considered here would (had
it been found to have $\theta = 90^\circ$) have had a diameter
increase of 70\% along the line of sight. Hence $L_{\rm obs}/D_{\rm
obs}$ would have changed from 6.6 to 3.9, and the corresponding
sparse sampling error from $1.9^\circ$ to $3.4^\circ$. From Table~2
we see that this 70\% increase has no effect, but with another factor of
two those errors would become dominating.

Errors perpendicular to the line of sight are typically a fraction of
an arcsec, or about two orders of magnitude less than those along the
line of sight. Those can be completely ignored. A side-effect of this
``orders of magnitude difference'' is that thin filaments with {\it
small} inclination angles will appear observationally as thin filaments
(strings), while thin filaments with {\it large} inclination angles
will appear to have cross-sections which are elongated along the line
of sight (ribbons).
Observational determination of the magnitude of this ``ribbon effect''
will make it possible to determine the {\it actual} RMS of the
non-Hubble flow velocities
at $z=3$ (the 200~km~s$^{-1}$ we used in this example is, as mentioned above,
to be regarded as an upper limit).

The 20~Mpc filament considered above remains fully defined when we
include expected random errors. However, we see from Fig.~6 that a
similar filament defined by only 4 LEGOs would move close to the
curve marking the $10^\circ$ sparse sampling error. Therefore,
because of the expected magnitude of the errors caused by random motion,
it will in general not be possible to determine inclination angles with
sufficient precision for filaments significantly shorter than 15~Mpc.
Filaments significantly larger than 50~Mpc are still prohibitively
difficult to identify for observational reasons, so at present an
optimal search for filaments at $z=3$ must target filaments of
comoving length $15-50$~Mpc. The optimal survey box has sides
of length $50-60$~Mpc.

\subsubsection{Systematic peculiar velocities}
The final point we shall consider is the effect of bulk flow in
the survey box. First we address the scale dependence.
Any bulk motion on the same scale as the filaments themselves, or
larger, will simply result in a translation of an entire filament
and hence have no influence on the distribution of angles. Bulk motion
on scales much smaller than the filaments will appear as random
motion (treated in the previous section). Therefore, only bulk
motion in the regime between those two limits (several, but not all,
LEGOs in a single filament) is relevant.

It is at present not easy to predict precisely what sort of bulk
motion will be present on this scale. In the local universe we see
large scale streaming towards massive dark matter halos, but since 
this is the result of acceleration over a Hubble time one would expect 
similar streaming at higher redshifts to have smaller amplitude.
In simulations of the
early universe two types of flows are seen. First it is often seen
that protogalaxies in the filaments stream along the filaments
towards the nodes, second it is seen that nodes are attracted to
each other. The ``streaming towards nodes'' will have the effect of
making the filaments appear longer than they are (stretching), the
movement of nodes towards each other will have the opposite effect
(squeezing). To first order one might expect that the two
effects cancel each other, but it is equally possible that one of the
two dominates. If the latter is the case, it will cause a systematic
error on the determination of $F$.

In the previous sections we have outlined how to determine $F$ from a
set of data. Those data could be observational, but they could equally
well be taken from a simulation. The only way to address the possible
effect of intermediary scale bulk motion is to apply our method to
sets of filaments in simulated volumes. We shall return to the results
of this work in a separate paper.

\section{Summary and conclusions}\label{summary}
 A significant number of $8-10$ meter class telescopes became operational
during the past decade, thereby moving several cosmological tests from the
domain of speculation into the domain of observation. The currently most
successful test relies heavily upon the assumption that supernovae of type
Ia are standard candles. Other proposed tests, based on e.g.~distances
between Lyman Break galaxies, seek to exploit an expected standard rod.

In this paper we have presented a detailed discussion of a test which
relies on neither standard candles nor standard rods. The test makes use
of only the requirement of isotropy, and of the prediction from numerical
simulations of the high redshift universe that it has a filamentary
structure. Direct observational evidence that such filaments may indeed be
found was presented by M{\o}ller \& Fynbo (2001) who also proposed three
cosmological tests based on the existence of such filaments. Here we have
in detail investigated and described the implementation of the first of
these tests. We have used Monte Carlo simulations to determine the
accuracy with which one may obtain values of $\Omega_{\rm m}$ and
$\Omega_\Lambda$, and we have described how additional statistical errors
and observational effects will affect the results. Guided by the detailed
Monte Carlo simulations, and taking into account the capabilities of
current instrumentation, we have considered what would be the optimal
survey parameters for an observing campaign aimed at an implementation of
this test. 

We find that the constraints this test can set are mostly on
$\Omega_\Lambda$, or in other words the limiting curves in the usual
($\Omega_{\rm m}$, $\Omega_\Lambda$) diagram are mostly horizontal. This
means that they intersect the probability curves from the $z < 1$ SN Ia
projects. Already with a sample of only 20 filaments the SN Ia results
will be intersected, but with a sample of 50 filaments the area of allowed
values in the ($\Omega_{\rm m}$, $\Omega_{\Lambda}$) diagram will be
significantly reduced. The optimal redshift range for the test is found to
be $z=1.5-4$. In the range $z=2-4$ filaments can conveniently be detected
using deep searches for faint Ly$\alpha$ emitters (LEGOs). Each filament
to be used in this test must have an observed length-to-diameter ratio of
no less than 3, and must be defined by no less than four objects.

We have considered the effect of random and systematic non-Hubble
flow velocities. There are no observational determinations of peculiar
velocities at high redshifts yet, so their impact is discussed
in qualitative terms mostly. A detailed analysis of those effects, which
are important as they set the minimum scale size of structures which
will be useful for the test, will be presented in a future paper. However,
using the local RMS value for peculiar velocities (which we consider a
realistic upper limit), we find that the optimal survey box is a cube
with sides of length $50-60$ comoving Mpc. 

Three surveys are currently underway which should produce data useful for
this test: i) the Large Area Lyman Alpha (LALA) survey (Rhoads et
al.~2000) that targets Ly$\alpha$ emitters at redshift $4.5$ and $5.7$,
ii) a survey conducted with the SUBARU telescope which has resulted in a
large sample of candidate Ly$\alpha$ emitters at $z=4.86$ (Ouchi et
al.~2002), and iii) in the opposite end of the optimal redshift window
there is a survey on the Nordic Optical Telescope (NOT) to map out a slice
of the $z=2$ universe in Ly$\alpha$ emission (M{\o}ller et al.~in prep).
The NOT survey started in 2001 and takes advantage of the excellent UV
capabilities of the NOT instrumentation which was demonstrated in similar
observations of fields containing $z=2$ host galaxies of Gamma Ray
Bursters (Fynbo et al.~2002). Currently there are no large area Ly$\alpha$
surveys targeting $z=3$, and there are no large area Ly$\alpha$ surveys
conducted on the ESO VLT. Smaller targeted programmes aimed at $z=3$
fields have been successfully conducted on the VLT (Fynbo et al.~in prep),
confirming that the necessary volume density of Ly$\alpha$ emitters will
indeed be found.

\section*{Acknowledgments}
 We are grateful to H.-J.~Gerber for interesting discussions, and for
helpful comments on an earlier version of this manuscript. MPE and MW
acknowledge support from the ESO Directors Discretionary Fund.

\appendix
\section{Cosmology with $\Lambda\neq 0$}\label{sec:relations}
We follow the treatment given in Longair (1998), which uses the metric in
the form:
\begin{small}
\begin{equation}
ds^2 = c^2dt^2-R^2(t)[dr^2+{\mathcal R}^2\sin(r/{\mathcal R})^2
(d\theta^2+\sin(\theta)^2d\phi^2)],
\nonumber
\end{equation}
\end{small}
where $\mathcal R$ is the radius of curvature of the universe, and $R$ is
the scale factor normalized to 1 at the present epoch. The radial
coordinate $r$ is the metric distance at time $t$, and it can be found as
a function of $z$ by integrating the differential equation

\begin{equation}\frac{dr}{dz} =
\frac{c}{H_0}\left(\left(1+z\right)^2\left(\Omega_{\rm m}z+1\right) -
\Omega_{\Lambda}z\left(z+2\right)\right)^{-1/2}, \end{equation} where
$\Omega_{\rm m}$ and
$\Omega_{\Lambda}$ are given by the present mean density $\rho_0$ and the
cosmological constant $\Lambda$, respectively:
\begin{equation}\Omega_{\rm m} = \frac{8\pi G\rho_0 c^2}{3H_0^2},\:\:\:\:\:
  \Omega_{\Lambda}=\frac{\Lambda c^2}{3H_0^2}.\end{equation}
The radius of curvature $\mathcal R$ can be determined from the relation
\begin{equation}\frac{1}{{\mathcal R}^2} = \frac{\Omega_{\rm
m}+\Omega_{\Lambda}-1}{c^2/H_0^2}.\end{equation}
Longair defines the distance measure
\begin{equation}
D = \left\{ \begin{array}{ll}
              {\mathcal R}\sin\left(r/{\mathcal R}\right)   &
         \mbox{if $1/{\mathcal R}^2>0$ } \\
              r   &
         \mbox{if $1/{\mathcal R}^2=0$} \\
              {\mathcal R}\sinh\left(r/{\mathcal R}\right) &
         \mbox{if $1/{\mathcal R}^2<0$ }
            \end{array}
    \right.
\end{equation}
from which the angular diameter distance $D_\mathrm{A}$ can be derived as
$D_\mathrm{A} = D/(1+z)$. The proper radial distance is determined from
the differential $\left(\frac{dr}{dz}\right)_\mathrm{prop}=\frac{dr}{dz}/(1+z)$.

We can now finally derive the two functions $f_L$ and $f_W$ introduced in
Sect.~2. As $W_{\rm true} = \phi D_A$ the function $f_W$ is simply given by
$H_0 D_A$. Similarly, the proper length $L_{\rm true}$ of a box defined by two
redshifts $z_\mathrm{min}$ and $z_\mathrm{max}$
($\delta z = z_\mathrm{max}-z_\mathrm{min} \ll
(z_\mathrm{max}+z_\mathrm{min})/2$) is given by
$\left(\frac{dr}{dz}\right)_\mathrm{prop} \delta z$, hence
$f_L = H_0 \left(\frac{dr}{dz}\right)_\mathrm{prop}$.


\begin{thebibliography}{99}
\bibitem{A}Alcock, C. \& Paczy\'nski, B., 1979, Nature, 281, 358
\bibitem{B2001} Branchini, E., Freudling, W., Da Costa, L.N., et al.,
2001, MNRAS, 326, 1191
\bibitem{F}Felten, J.E., Isaacman, R., 1986, Rev. Mod. Phys., 58, 689
\bibitem{FMT2001}Fynbo, J.U., M\o ller P., Thomsen, B., 2001, A\&A, 374,
443
\bibitem{FMT2002}Fynbo, J.U., M\o ller P., Thomsen, B., et al., 2002, submitted
to A\&A
\bibitem{J}Jaffe, A.H, Ade, P.A.R., Balbi, A., et al., 2000, astro-ph/0007333
\bibitem{Kl}Klypin, A.A. \& Shandarin, S.F., 1983, MNRAS, 204, 891
\bibitem{L}Longair, M.S., 1998, chapter 7 in: "Galaxy Formation",
A\&A Library (eds. I. Appenzeller et al.), Springer Verlag
\bibitem{Mc}McDonald, P. \& Miralda-Escud\'e, J., 1999, ApJ, 518, 24
\bibitem{MW1998}M\o ller, P. \& Warren, S.J., 1998, MNRAS, 299, 661
\bibitem{M}M\o ller, P. \& Fynbo, J.U., 2001, A\&A, 372, L57
\bibitem{O}Ouchi, M., Shimasaku, K., Furusawa, H., et al., 2002,
submitted to ApJ (astro-ph/0202204)
\bibitem{Pe}Perlmutter, S., Aldering, G., Goldhaber, G., et al., 1999, ApJ,
517, 565
\bibitem{Pr}Press, W.H, Flannery, B.P, Teukolsky, S.A, Vetterling, W.T., 1989,
``Numerical Recipes'',
Cambridge University Press
\bibitem{Ra}Rauch, M., Haehnelt, M.G., Steinmetz, M., 1997, ApJ, 481, 601
\bibitem{LALA}Rhoads, J.E., Malhotra, S., Dey, A., et al., 2000, ApJL,
545, 85
\bibitem{R}Riess, A.G., Filippenko, A.V., Challis, P., et al.,1998, ApJ, 116,
1009
\bibitem{Ro}Roukema, B.F., 2001, A\&A, 369, 729
\bibitem{RR}Rowan-Robinson, M., 2002, MNRAS, 332, 352
\bibitem{Sandage}Sandage, A., 1961, ApJ, 133, 355
\bibitem{SS}Simonsen, J.T., Hannestad, S., 1999, A\&A, 351, 1
\bibitem{Wh}White, S.D.M., Frenk, C.S., Davis, M., Efstathiou, G., 1987, ApJ,
313, 505
\end{thebibliography}
\end{document}